\documentclass[conference]{IEEEtran}
\IEEEoverridecommandlockouts
\usepackage{cite}
\usepackage{amsmath,amssymb,amsfonts}
\usepackage{algorithmic}
\usepackage{graphicx}
\usepackage{textcomp}
\usepackage{xcolor}
\definecolor{chimgreen}{HTML}{168638}
\newcommand{\best}[1]{\textcolor{chimgreen}{\textbf{#1}}}
\usepackage{colortbl}
\definecolor{soaAccel}{HTML}{B7D2E7}  
\definecolor{soaSoc}{HTML}{b9dac3}    
\usepackage{makecell}
\usepackage{pifont}
\newcommand{\cmark}{\ding{51}}
\newcommand{\xmark}{\ding{55}}
\usepackage{glossaries}
\usepackage{orcidlink}
\usepackage{siunitx}
\usepackage[nameinlink]{cleveref}
\Crefname{figure}{Fig.}{Figs.}
\crefname{figure}{Fig.}{Figs.}
\usepackage[position=b]{subcaption}
\usepackage[font=scriptsize]{caption}

\DeclareSIUnit\ops{OPS}
\DeclareSIUnit\bs{b/s}
\DeclareSIUnit\ge{GE}

\newcommand{\ResultSoaChimeraPeakEffRatioSoc}{1.37}
\newcommand{\ResultSoaChimeraAreaEffRatioSoc}{100}
\newcommand{\ResultSoaChimeraAreaEffRatioAcc}{1.8}
\newcommand{\ResultSoaAyakaDieArea}{10.76}
\newcommand{\ResultSoaEvaDieArea}{16}
\newcommand{\ResultSoaFanDieArea}{5.8}
\newcommand{\ResultSoaTinyVersDieArea}{6.25}
\newcommand{\ResultSoaStmDieArea}{5.52}
\newcommand{\ResultSoaChimeraDieArea}{3.19}
\newcommand{\ResultSoaAyakaPeakPerfMin}{170}
\newcommand{\ResultSoaAyakaPeakPerfMax}{6530}
\newcommand{\ResultSoaEvaPeakPerf}{1544}
\newcommand{\ResultSoaFanPeakPerf}{920}
\newcommand{\ResultSoaTinyVersPeakPerf}{17.6}
\newcommand{\ResultSoaChimeraPeakPerf}{896}
\newcommand{\ResultSoaAyakaPeakEffMin}{2.22}
\newcommand{\ResultSoaAyakaPeakEffMax}{49.7}
\newcommand{\ResultSoaEvaPeakEff}{0.71}
\newcommand{\ResultSoaFanPeakEffMin}{4.46}
\newcommand{\ResultSoaFanPeakEffMax}{12.52}
\newcommand{\ResultSoaTinyVersPeakEff}{2.47}
\newcommand{\ResultSoaStmPeakEff}{3.25}
\newcommand{\ResultSoaStmPeakEffScaledDeficit}{27}
\newcommand{\ResultSoaChimeraPeakEff}{3.1}
\newcommand{\ResultSoaAyakaAreaEffMin}{15.8}
\newcommand{\ResultSoaAyakaAreaEffMax}{607}
\newcommand{\ResultSoaEvaAreaEff}{96.5}
\newcommand{\ResultSoaFanAreaEff}{158.62}
\newcommand{\ResultSoaTinyVersAreaEff}{2.82}
\newcommand{\ResultSoaChimeraAreaEff}{281}

\makeatletter
\newcommand{\insertnewlines}[1]{%
  \noindent\mbox{}%
  \@tempcnta=#1\relax
  \loop\ifnum\@tempcnta>0
    \\
    \advance\@tempcnta\m@ne
  \repeat
}
\makeatother

\newacronym[plural=WANs, firstplural={Wide Area Networks (WANs)}]{wan}{WAN}{Wide Area Network}
\newacronym[plural=WSNs, firstplural={Wireless Sensor Networks (WSNs)}]{wsn}{WSN}{Wireless Sensor Network}
\newacronym{simd}{SIMD}{Single Instruction Multiple Data}
\newacronym{os}{OS}{Operating System}
\newacronym{ble}{BLE}{Bluetooth Low-Energy}
\newacronym{wifi}{Wi-FI}{Wireless Fidelity}
\newacronym[plural=DVS, firstplural={Dynamic Vision Sensors (DVS)}]{dvs}{DVS}{Dynamic Vision Sensor}
\newacronym{ptz}{PTZ}{Pan-Tilt Unit}

\newacronym[plural=FLLs,firstplural=Frequency Locked Loops (FLLs)]{fll}{FLL}{Frequency Locked Loop}
\newacronym{dram}{DRAM}{Dynamic Random Access Memory}
\newacronym{fpu}{FPU}{Floating Point Unit}
\newacronym{fpss}{FPSS}{Floating Point Subsystem}
\newacronym{frep}{FREP}{Floating Point Repetition}
\newacronym{dma}{DMA}{Direct Memory Access}
\newacronym{ssr}{SSR}{Stream Semantic Register}
\newacronym{issr}{ISSR}{Indirection Stream Semantic Register}
\newacronym[plural=LUTs, firstplural={Lookup Tables (LUTs)}]{lut}{LUT}{Lookup Table}
\newacronym[plural=FPGAs, firstplural={Field Programmable Gate Arrays (FPGAs)}]{fpga}{FPGA}{Field Programmable Gate Array}
\newacronym{dsp}{DSP}{Digital Signal Processing}
\newacronym[plural=MCUs, firstplural={Microcontroller Units (MCUs)}]{mcu}{MCU}{Microcontroller Unit}
\newacronym[plural=AI-MCUs, firstplural={AI-accelerated Microcontroller Units (AI-MCUs)}]{ai-mcu}{AI-MCU}{AI-accelerated Microcontroller Unit}
\newacronym{spi}{SPI}{Serial Peripheral Interface}
\newacronym{cpi}{CPI}{Camera Parallel Interface}
\newacronym{rf}{RF}{Register File}
\newacronym{fifo}{FIFO}{First-In First-Out Queue}
\newacronym{uart}{UART}{Universal Asynchronous Receiver-Transmitter}
\newacronym{raw}{RAW}{Read After Write}
\newacronym[plural=ISAs, firstplural={Instruction Set Architectures (ISAs)}]{isa}{ISA}{Instruction Set Architecture}
\newacronym{xbar}{XBAR}{crossbar}
\newacronym[firstplural=Scratch-Pad Memories (SPMs)]{spm}{SPM}{Scratch-Pad Memory}
\newacronym{ppa}{PPA}{Power Performance Area}
\newacronym{ipi}{IPI}{Inter-Processor Interrupt}
\newacronym[firstplural=Software-Generated Interrupts (SGIs)]{sgi}{SGI}{Software-Generated Interrupt}
\newacronym[firstplural=Processing Elements (PEs)]{pe}{PE}{Processing Element}
\newacronym{tcdm}{TCDM}{Tightly-Coupled Data Memory}
\newacronym{lsu}{LSU}{Load-Store Unit}
\newacronym{icache}{I\$}{Instruction Cache}
\newacronym{dcache}{D\$}{Data Cache}
\newacronym{wfi}{WFI}{Wait For Interrupt}
\newacronym{gp}{GP}{General-Purpose}
\newacronym{gpc}{GPC}{GPU Processing Cluster}
\newacronym{cpu}{CPU}{Central Processing Unit}
\newacronym{gpu}{GPU}{Graphics Processing Unit}
\newacronym{llc}{LLC}{Last-Level Cache}
\newacronym{sm}{SM}{Streaming Multiprocessor}
\newacronym[firstplural=Networks on Chip (NoCs)]{noc}{NoC}{Network on Chip}
\newacronym[firstplural=Virtual Channels (VCs)]{vc}{VC}{Virtual Channel}
\newacronym[firstplural=Network Interfaces (NIs)]{ni}{NI}{Network Interface}

\newacronym{dfg}{DFG}{Data Flow Graph}
\newacronym{lcg}{LCG}{Linear Congruential Generator}
\newacronym{prn}{PRN}{Pseudo-Random Number}

\newacronym{ste}{STE}{Straight-Through-Estimator}

\newacronym[plural=PTUs, firstplural={Pan-Tilt Units}]{ptu}{PTU}{Pan-Tilt Unit}
\newacronym{mdf}{MDF}{Medium-density fibreboard}
\newacronym{cvat}{CVAT}{Computer Vision Annotation Tool}
\newacronym{coco}{COCO}{Common Objects in Context}
\newacronym{soa}{SoA}{State of the Art}
\newacronym{sf}{SF}{Sensor Fusion}

\newacronym{tac}{TAC}{Transformer Acceleration Cluster}
\newacronym{e2e}{E2E}{End-to-End}

\newacronym{dl}{DL}{Deep Learning}
\newacronym{bn}{BN}{Batch Normalization}
\newacronym{FGSM}{FBK}{Fast Gradient Sign Method}
\newacronym{lr}{LR}{Learning Rate}
\newacronym{sgd}{SGD}{Stochastic Gradient Descent}
\newacronym{gd}{GD}{Gradient Descent}
\newacronym{llm}{LLM}{Large Language Model}

\newacronym{sta}{STA}{Static Timing Analysis}

\newacronym[plural=GPIOs, firstplural={General Purpose Inupt Outputs (GPIOs)}]{gpio}{GPIO}{General Purpose Input Output}
\newacronym[plural=LDOs, firstplural={Low Dropout Regulators (LDOs)}]{ldo}{LDO}{Low Dropout Regulator}

\newacronym{inq}{INQ}{Incremental Network Quantization}

\newacronym{cv}{CV}{Computer Vision}
\newacronym{EoT}{EoT}{Expectation over Transformation}
\newacronym{RPN}{RPN}{Region Proposal Network}
\newacronym{TV}{TV}{Total Variation}
\newacronym{NPS}{NPS}{Non-Printability Score}
\newacronym{STN}{STN}{Spatial Transformer Network}
\newacronym{MTCNN}{MTCNN}{Multi-Task Convolutional Neural Network}
\newacronym{YOLO}{YOLO}{You Only Look Once}
\newacronym{SSD}{SSD}{Single Shot Detector}
\newacronym{SOTA}{SOTA}{State of the Art}
\newacronym{NMS}{NMS}{Non-Maximum Suppression}
\newacronym{ic}{IC}{Integrated Circuit}
\newacronym{tcxo}{TCXO}{Temperature Controlled Crystal Oscillator}
\newacronym{jtag}{JTAG}{Joint Test Action Group industry standard}
\newacronym{swd}{SWD}{Serial Wire Debug}
\newacronym{sdio}{SDIO}{Serial Data Input Output}

\newacronym[plural=PCBs, firstplural={Printed Circuit Boards (PCB)}]{pcb}{PCB}{Printed Circuit Board}
\newacronym[plural=ASICs, firstplural={Application Specific Integrated Circuits}]{asic}{ASIC}{Application Specific Integrated Circuit}

\newacronym[plural=BNNs, firstplural={Binary Neural Networks (BNNs)}]{bnn}{BNN}{Binary Neural Network}
\newacronym[plural=NNs, firstplural={Neural Networks}]{nn}{NN}{Neural Network (NNs)}
\newacronym[plural=SCMs, firstplural={Standard Cell Memories (SCMs)}]{scm}{SCM}{Standard Cell Memory}
\newacronym{ann}{ANN}{Artificial Neural Networks}
\newacronym{ml}{ML}{Machine Learning}
\newacronym{ai}{AI}{Artificial Intelligence}
\newacronym{iot}{IoT}{Internet of Things}
\newacronym{fft}{FFT}{Fast Fourier Transform}
\newacronym[plural=OCUs, firstplural={Output Channel Compute Units (OCUs)}]{ocu}{OCU}{Output Channel Compute Unit}
\newacronym{alu}{ALU}{Arithmetic Logic Unit}
\newacronym{mac}{MAC}{Multiply-Accumulate}
\newacronym[firstplural={systems-on-chip (SoCs)}]{soc}{SoC}{system-on-chip}
\newacronym[firstplural={multi-processor systems-on-chip (MPSoCs)}]{mpsoc}{MPSoC}{multi-processor system-on-chip}
\newacronym{dut}{DUT}{Device Under Test}
\newacronym{qos}{QoS}{Quality of Service}

\newacronym{PGD}{PGD}{Projected Gradient Descend}
\newacronym{CW}{CW}{Carlini-Wagner}
\newacronym{OD}{OD}{Object Detection}

\newacronym{rrf}{RRF}{RADAR Repetition Frequency}
\newacronym{nlp}{NLP}{Natural Language Processing}
\newacronym[firstplural={Vision Transformers (ViTs)}]{vit}{ViT}{Vision Transformer}
\newacronym{qam}{QAM}{Quadrature Amplitude Modulation}
\newacronym{rri}{RRI}{RADAR Repetition Interval}
\newacronym{radar}{RADAR}{Radio Detection and Ranging}
\newacronym{loocv}{LOOCV}{Leave-one-out cross validation}

\newacronym{bsp}{BSP}{Board Support Package}
\newacronym{ttn}{TTN}{The Things Network}
\newacronym{wip}{WIP}{Work in Progress}
\newacronym{json}{JSON}{JavaScript Object Notation}
\newacronym{qat}{QAT}{Quantization-Aware Training}

\newacronym{cls}{CLS}{Classification Error}
\newacronym{loc}{LOC}{Localization Error}
\newacronym{bkgd}{BKGD}{Background Error}
\newacronym{roc}{ROC}{Receiver Operating Characteristic}
\newacronym{frr}{FRR}{False Rejection Rate}
\newacronym{eer}{EER}{Equal Error Rate}
\newacronym{snr}{SNR}{Signal-to-Noise Ratio}
\newacronym{flop}{FLOP}{Floating-Point Operation}
\newacronym{fp}{FP}{Floating-Point}
\newacronym{fps}{FPS}{Frames Per Second}
\newacronym{oi}{OI}{Operational Intensity}
\newacronym[first={IPC (Instructions per Cycle)}]{ipc}{IPC}{Instructions per Cycle}

\newacronym{gsc}{GSC}{Google Speech Commands}
\newacronym{mswc}{MSWC}{Multilingual Spoken Words Corpus}
\newacronym{demand}{DEMAND}{Diverse Environments Multichannel Acoustic Noise Database}

\newacronym[plural=SNNs, firstplural={Spiking Neural Networks (SNNs)}]{snn}{SNN}{Spiking Neural Network}
\newacronym[plural=DNNs, firstplural={Deep Neural Networks (DNNs)}]{dnn}{DNN}{Deep Neural Network}
\newacronym[plural=TCNs,firstplural=Temporal Convolutional Networks]{tcn}{TCN}{Temporal Convolutional Network}
\newacronym[plural=CNNs,firstplural=Convolutional Neural Networks (CNNs)]{cnn}{CNN}{Convolutional Neural Network}
\newacronym[plural=TNNs,firstplural=Ternarized Neural Networks]{tnn}{TNN}{Ternarized Neural Network}
\newacronym{ds-cnn}{DS-CNN}{Depthwise Separable Convolutional Neural Network}
\newacronym{rnn}{RNN}{Recurrent Neural Network}
\newacronym{gcn}{GCN}{Graph Convolutional Network}
\newacronym{mhsa}{MHSA}{Multi-Head Self Attention}
\newacronym{mha}{MHA}{Multi-Head Attention}
\newacronym{crnn}{CRNN}{Convolutional Recurrent Neural Network}
\newacronym{clca}{CLCA}{Convolutional Linear Cross-Attention}

\newacronym{bf}{BF}{Beamforming}
\newacronym{anc}{ANC}{Active Noise Cancellation}
\newacronym{agc}{AGC}{Automatic Gain Control}
\newacronym{se}{SE}{Speech Enhancement}
\newacronym{mct}{MCT}{Multi-Condition Training}
\newacronym{mcta}{MCTA}{Multi-Condition Training \& Adaptation}
\newacronym{pcen}{PCEN}{Per-Channel Energy Normalization}
\newacronym{mfcc}{MFCC}{Mel-Frequency Cepstral Coefficient}
\newacronym{asr}{ASR}{Automated Speech Recognition}
\newacronym{kws}{KWS}{Keyword Spotting}
\newacronym{odl}{ODL}{On-Device Learning}

\newacronym{nl-kws}{NL-KWS}{Noiseless Keyword Spotting}
\newacronym{na-kws}{NA-KWS}{Noise-Aware Keyword Spotting}
\newacronym{odda}{ODDA}{On-Device Domain Adaptation}
\newacronym{hpm}{HPM}{High-Performance Mode}
\newacronym{lpm}{LPM}{Low-Power Mode}

\hypersetup{hidelinks}

\def\BibTeX{{\rm B\kern-.05em{\sc i\kern-.025em b}\kern-.08em
    T\kern-.1667em\lower.7ex\hbox{E}\kern-.125emX}}
\begin{document}

\title{CHIMERA: A Flexible and Scalable 3.1 TOPS/W AI-MCU with Transformer Accelerator and 563 Gb/s Shared-L2 Memory Subsystem with QoS Guarantees}

\ifdefined\blindreview
\author{\centering{\textit{Authors omitted for blind review.}}}
\else
\author{
    \IEEEauthorblockN{Lorenzo Leone\textsuperscript{*}\orcidlink{0009-0000-3976-847X}, Philip Wiese\textsuperscript{*}\orcidlink{0009-0001-7214-2150}, Gamze Islamoglu\textsuperscript{*}\orcidlink{0000-0002-5129-1691},
    Michael Rogenmoser\textsuperscript{*}\orcidlink{0000-0003-4622-4862}\\ Davide Rossi\textsuperscript{\dag\ddag}\orcidlink{0000-0002-0651-5393}, Francesco Conti\textsuperscript{\dag}\orcidlink{0000-0002-7924-933X}, Luca Benini\textsuperscript{*\dag}\orcidlink{0000-0001-8068-3806}}
    \IEEEauthorblockA{\textsuperscript{*}\textit{ETH Zürich, Zürich, Switzerland}, \textsuperscript{\dag}\textit{Università di Bologna, Bologna, Italy}, \textsuperscript{\ddag}\textit{Chips-IT, Pavia, Italy}}
    \textsuperscript{*}\{lleone, wiesep, gislamoglu, michaero, lbenini\}@iis.ee.ethz.ch, \textsuperscript{\dag}\{davide.rossi, f.conti\}@unibo.it
}
\fi

\maketitle

\begin{abstract}
    We present Chimera, a flexible and scalable \gls{mcu} designed to accelerate real-time inference of rapidly evolving transformer-based models at the ultra-low-power edge (hundred of \SI{}{\milli\watt}). The chip, implemented in \SI{22}{nm} FDX technology, integrates a transformer accelerator tightly coupled within a compute cluster featuring nine general-purpose RV32IMA cores. Scalability extends to the memory hierarchy through a novel L2 memory island subsystem, which enables data sharing across multiple clusters while delivering \SI{563}{\giga\bs} aggregate bandwidth. The L2 subsystem enforces quality-of-service guarantees for latency-critical traffic, achieving up to 16$\times$ latency reduction. Chimera achieves peak energy and area efficiencies of \SI{\ResultSoaChimeraPeakEff}{\tera\ops/\watt} and \SI{\ResultSoaChimeraAreaEff}{\giga\ops/\milli\metre\squared}, demonstrating \ResultSoaChimeraPeakEffRatioSoc$\times$ higher energy efficiency and up to \ResultSoaChimeraAreaEffRatioSoc$\times$ higher area efficiency compared to \gls{soa} SoCs. Compared to \gls{soa} standalone accelerators, Chimera achieves comparable energy efficiency and up to \ResultSoaChimeraAreaEffRatioAcc$\times$ higher area efficiency.
\end{abstract}
\vspace{-0em}

\begin{IEEEkeywords}
    Transformers, Edge-AI, MCU, QoS, Attention
\end{IEEEkeywords}

\section{Introduction}
\begin{figure}[t]
    \centering
    \begin{subfigure}[t]{\columnwidth}
        \centering
        \includegraphics[width=\textwidth]{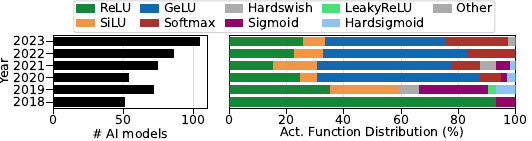}
        \phantomcaption
        \label{fig:intro_barchart}
    \end{subfigure}
    \vspace{-1.8em}

    \begin{subfigure}[t]{\columnwidth}
        \centering
        \includegraphics[width=\textwidth]{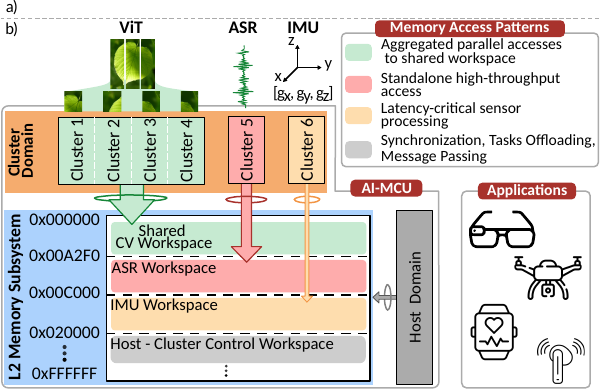}
        \phantomcaption
        \label{fig:application}
    \end{subfigure}
    \vspace{-1em}
    \caption{(a) Growth and diversification of \gls{ai} models \cite{standford_ai25} and activation functions \cite{andri23} over time. b) Multi-cluster workload execution patterns and their impact on shared L2 memory.}
    \label{fig:intro}
    \vspace{-1.2em}
\end{figure}

The increasing diffusion of \gls{ai} workloads, such as \gls{nlp} and speech recognition, in edge and TinyML systems drives the need for high-throughput \glspl{ai-mcu} supporting real-time-constrained execution under strict power and area budgets (from tens to a few hundred \SI{}{\milli\watt} and tens of \SI{}{\milli\metre\squared}) \cite{silvano25}.
The rapid evolution of \gls{ai} models toward attention-based \glspl{dnn} (\Cref{fig:intro_barchart}) demands flexibility, motivating system-level co-design of tightly coupled clusters of programmable processors and specialized acceleration engines sharing L1 memory \cite{verhelst25}.
At the same time, the increasing heterogeneity and scale of TinyML workloads \cite{aminabadi22} are driving a shift toward multi-cluster architectures, which in turn places significant pressure on the L2 memory, shared among all clusters. Sustaining multiple clusters therefore requires high aggregate L2 bandwidth (\Cref{fig:application}), minimizing inter-cluster contention \cite{gholami24, dagli24}.
Moreover, in \gls{ai-mcu}, a host core orchestrates synchronization and message passing across clusters, generating latency-critical traffic that requires fast and predictable service. This makes both average and worst-case access latency critical, motivating \gls{qos} support in the L2 memory subsystem \cite{bechtel24}.

We present Chimera%
\footnote{\vspace{-3em}
\ifdefined\blindreview
    https://hidden-for-double-blind-review.com
\else
    \url{https://github.com/pulp-platform/chimera/releases/tag/CONVOLVE-TO}
\fi
}, a flexible \gls{ai-mcu} that addresses these challenges through three key innovations:
\textbf{(A)} an energy-efficient \gls{tac} integrating a transformer accelerator tightly coupled with fully programmable RV32IMA cores, enabling flexibility and adaptability to rapidly evolving models (\Cref{fig:intro_barchart}), achieving up to \SI{\ResultSoaChimeraPeakEff}{\tera\ops/\watt};
\textbf{(B)} a shared high-bandwidth AXI4-based L2 memory subsystem capable of delivering up to \SI{563}{\giga\bs} while mitigating inter-cluster contention, thereby enabling efficient multi-cluster workload parallelization;
\textbf{(C)} a QoS-aware L2 memory architecture enabling isolation of latency-critical accesses from concurrent high-throughput traffic, achieving a 34-cycle worst-case access latency and up to 16$\times$ latency reduction compared to conventional designs. By addressing memory bandwidth scalability and contention control, Chimera enables predictable, high-performance execution of heterogeneous TinyML workloads on a low-power AI-MCU.

\begin{figure*}[h]
    \centering
    \includegraphics[width=\textwidth]{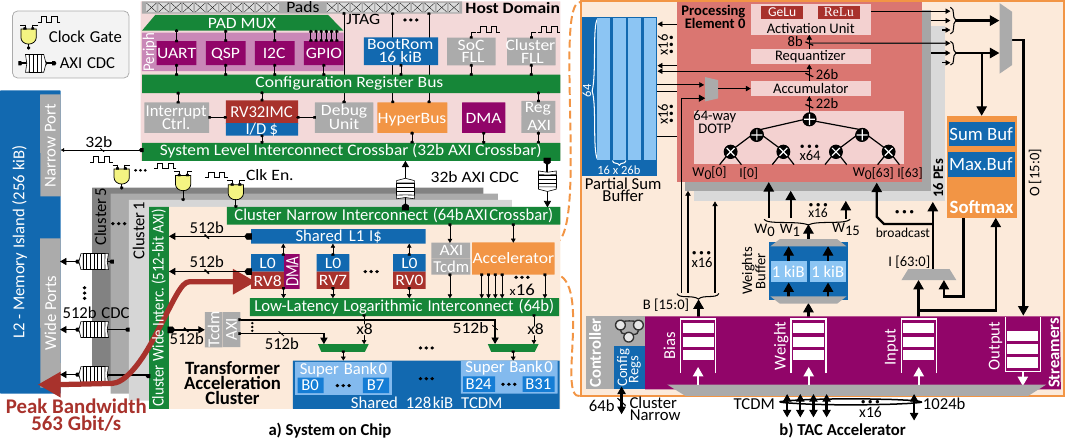}
    \begingroup
        \phantomsubcaption
        \label{fig:soc}
        \phantomsubcaption
        \label{fig:ita}
    \endgroup
    \vspace{-1.6em}
    \caption{(a) Architectural overview of the Chimera \gls{soc}. The clusters operate in a dedicated clock domain, while the host and memory island share a common clock. AXI clock-domain crossing modules ensure synchronization, and clock-gating cells at the cluster boundary enable software-controlled clock gating. (b) TAC architecture: configuration registers are programmed via the narrow AXI interface, while streamers handle TCDM data transfers. Weights (\texttt{W}) are stored in a \SI{2}{\kibi\byte} double-buffered memory, enabling overlap of computation and data movement. Input activations (\texttt{I}) are broadcast to all PEs, which compute 64-way dot products, producing 16 output elements per cycle (\texttt{O}). Softmax is computed on-the-fly for attention, while ReLU and GeLU are handled by the activation unit.}
    \label{fig:top_architecture}
    \vspace{-1em}
\end{figure*}

\section{Architecture}

\begin{figure}[t]
    \centering
    \includegraphics[width=\columnwidth]{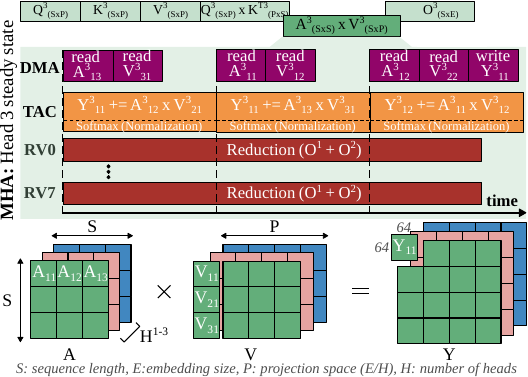}
    \caption{Scheduling of \gls{mha} on the TAC cluster. While the accelerator computes a tile, the DMA prepares data for the next tile, and the \gls{gp} cores reduce previously computed heads. The DMA can sustain computational throughput thanks to the memory island subsystem.}
    \label{fig:ita_scheduling}
    \vspace{-1em}
\end{figure}

Chimera is a multi-cluster \gls{soc} (\Cref{fig:soc}) integrating seven domains to address the heterogeneous demands of TinyML signal processing.
It includes a host domain, five heterogeneous clusters, and a shared L2 memory island. The host comprises an RV32IMC core responsible for system management and coordination, along with a rich peripheral subsystem including UART, I\textsuperscript{2}C, a HyperBus controller, and a \gls{dma} engine supporting data transfers between L3 memory and on-chip memories.

In this work, we focus on the \acrlong{tac} (\gls{tac}).
It includes eight RV32IMA cores sharing a \SI{128}{\kibi\byte} \gls{tcdm}, along with a \SI{4}{\kibi\byte} L1 I-cache.
A ninth core is dedicated to \gls{dma} management, orchestrating high-throughput transfers between the cluster and L2 memory via 512-bit ports, enabling efficient AXI4 burst transactions.
\begin{figure}[t]
    \centering
    \includegraphics[width=\columnwidth]{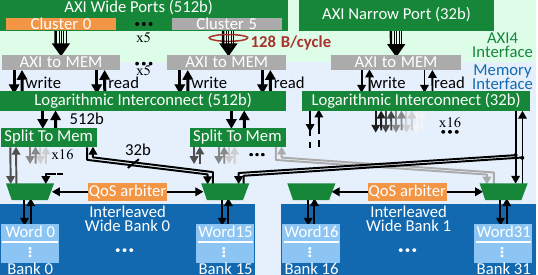}
    \vspace{-1em}
    \caption{L2 memory island architecture: forward arrows represent initiators, while backward arrows represent responses from target endpoints. The design features two interleaved wide banks delivering up to \SI{128}{\byte/cycle}, along with a \gls{qos}-aware arbitration policy for latency-critical accesses.}
    \label{fig:memisland}
    \vspace{-2em}
\end{figure}

Tightly coupled with the cores, the accelerator (\Cref{fig:ita}) supports GEMM and attention mechanism using 8-bit integer quantization with minimal accuracy loss \cite{islamoglu23}.
The accelerator comprises 16 \glspl{pe}, each operating on 8-bit weights (\texttt{W}) and activations (\texttt{I}), and computing a 64-way dot product per cycle, resulting in a peak throughput of \SI{2048}{op/cycle}.
Data is supplied through three streamers for inputs (\texttt{I}), weights (\texttt{W}), and biases (\texttt{B}), while a fourth streamer handles output write-back (\texttt{O}), each providing up to \SI{128}{\byte/cycle}.
To sustain this peak fetch bandwidth, the accelerator connects to the \gls{tcdm} interconnect via 16 64-bit master ports. The accelerator also integrates an activation unit (\SI{8.8}{\kilo\ge}) in each \gls{pe}, as well as a softmax engine (\SI{44}{\kilo\ge}) with a peak throughput of 64 softmax/cycle, operating concurrently with the \glspl{pe} during attention (\Cref{fig:ita_scheduling}).

\begin{figure}[t]
    \centering
    \includegraphics[width=\columnwidth]{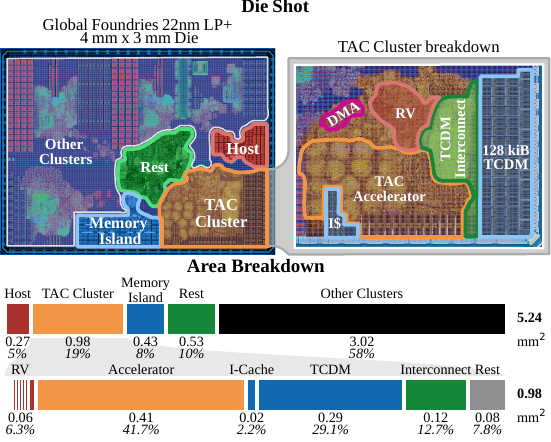}
    \vspace{-1.6em}
    \caption{Annotated chip micrograph and area breakdown. The overall die area is \SI{12}{\milli\metre\squared}. The silicon area evaluated in this work is \SI{3.19}{\milli\metre\squared} at 60\% logic area utilization.}
    \label{fig:die}
    \vspace{-0.5em}
\end{figure}

\begin{figure}[t]
    \centering
    \begin{subfigure}{\columnwidth}
        \centering
        \includegraphics[width=\textwidth]{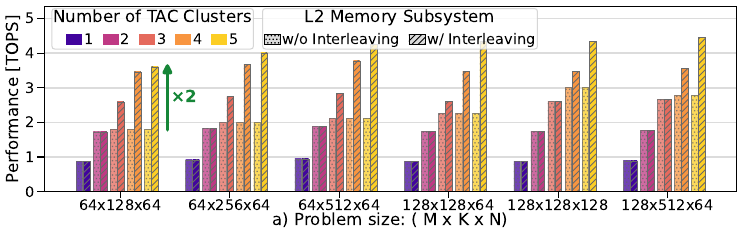}
        \vspace{-2em}
        \phantomcaption
        \label{fig:memisland_speedup}
    \end{subfigure}
    \vspace{0em}

    \begin{subfigure}{\columnwidth}
        \centering
        \includegraphics[width=\textwidth]{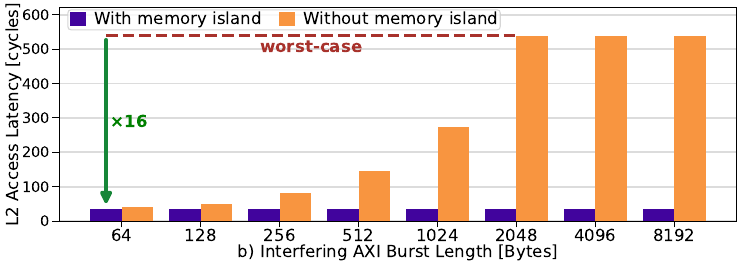}
        \phantomcaption
        \label{fig:qos_latency}
    \end{subfigure}
    \vspace{-2.5em}
    \caption{(a) Simulated performance for different MATMUL sizes across multi-cluster configurations, evaluated with and without the L2 interleaved scheme. (b) Measured average L2 narrow (32-bit) access latency, under concurrent high-throughput data transfers with varying AXI4 burst lengths.}
    \label{fig:memisland_experiments}
    \vspace{-1.5em}
\end{figure}

To support efficient data sharing and sustain high aggregate bandwidth across multiple clusters, Chimera features a shared \SI{256}{\kibi\byte} L2 memory island (\Cref{fig:memisland}) with heterogeneous interfaces: 512-bit AXI4 wide interfaces for high-throughput traffic and a 32-bit AXI4 narrow interface for latency-critical messages.
The wide interfaces deliver a total read/write bandwidth of \SI{128}{\byte/cycle} per port. To sustain this bandwidth under parallel accesses, the L2 is organized into two interleaved wide banks, each \SI{128}{\kibi\byte}, mitigating access conflicts and approaching the peak physical bandwidth.

However, under sustained high-throughput traffic, latency-critical messages may experience degraded \gls{qos}.
To address this, the L2 supports arbitration policies including fixed priority for narrow accesses, which is effective when narrow traffic is regulated at system level, and a bounded-priority scheme to prevent starvation of wide accesses under continuous contention. This ensures low-latency service (34-cycle worst-case) for inter-cluster and host-to-cluster control traffic while maintaining high throughput for data-intensive workloads.

\section{Results}
Fabricated in \textsc{GlobalFoundries' (GF)} \SI{22}{\nano\metre} \textsc{LP+} technology, Chimera is designed for energy-efficient transformer-based workloads. \Cref{fig:die} shows the annotated die micrograph. The chip occupies \SI{12}{\milli\metre\squared}, of which the subsystem presented in this work accounts for \SI{3.19}{\milli\metre\squared} at 60\% logic utilization.

To evaluate the impact of the memory island's aggregated bandwidth and interleaving scheme, matrix multiplication kernels were simulated by scaling the number of \gls{tac} clusters. As shown in \Cref{fig:memisland_speedup}, beyond two active clusters, a baseline \gls{soc} without the proposed L2 subsystem is bottlenecked by inter-cluster access conflicts in the shared memory. In contrast, the proposed interleaving scheme mitigates these conflicts, enabling higher effective bandwidth than the baseline despite identical physical bandwidth, sustaining the increased throughput demand, and  achieving up to 2$\times$ higher performance.

To assess the ability of the memory island to provide predictable service for latency-critical accesses under contention, \gls{qos} is evaluated by issuing 20,000 32-bit L2-to-L1 reads from the RV32IMC host core through the narrow interface, while the cluster \gls{dma} concurrently generates AXI burst reads targeting the same memory region.
As shown in \Cref{fig:qos_latency}, the baseline L2 architecture exhibits significant and burst-length-dependent latency inflation, failing to provide predictable latency. In contrast, Chimera maintains bounded and predictable host access latency under intensive high-throughput traffic from the \gls{tac} cluster, achieving up to a 16$\times$ latency reduction, confirming the effectiveness of the proposed arbitration policy.

\begin{figure}[t]
    \centering
    \includegraphics[width=\columnwidth]{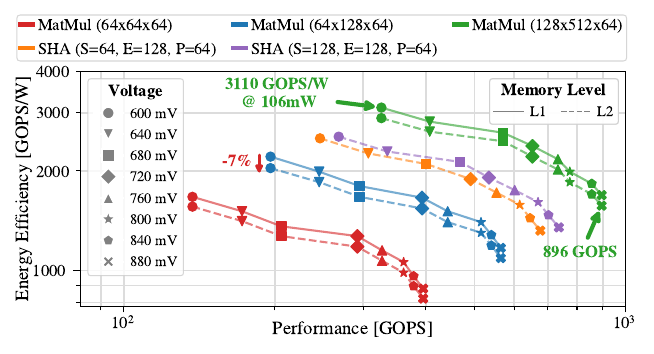}
    \vspace{-1.6em}
    \caption{Measured energy efficiency and performance for workloads executed from L1 and L2 using double buffering. Inactive clusters are clock-gated.}
    \label{fig:energy}
    \vspace{-1.5em}
\end{figure}

\begin{figure}[t]
    \centering
    \includegraphics[width=\columnwidth]{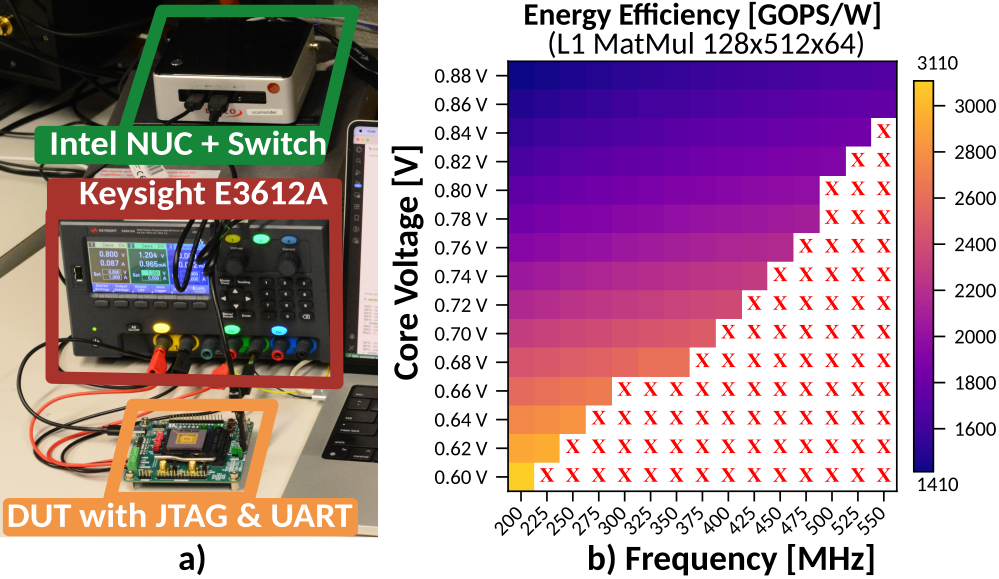}

    \begingroup
        \phantomsubcaption
        \label{fig:setup}
        \phantomsubcaption
        \label{fig:shmoo}
    \endgroup

    \caption{a) Measurement setup: the \gls{dut} si controlled via JTAG and UART, while a programmable power supply provides and measures the \gls{soc} supply voltage and current. b) Shmoo plot for a $128 \times 512 \times 64$ MATMUL.}
    \label{fig:setup_shmoo}
    \vspace{-1em}
\end{figure}

\Cref{fig:energy} summarizes performance and energy efficiency based on the testing setup shown in \Cref{fig:setup_shmoo}. When executing matrix multiplication and single-head attention from L1, Chimera achieves a peak efficiency of \SI{3.1}{\tera\ops/\watt} (\SI{200}{\mega\hertz}, \SI{0.6}{\volt}).
When the same workloads are executed from L2, the efficiency degrades by only 7\%, demonstrating the effectiveness of the proposed memory subsystem. In the high-performance corner (\SI{550}{\mega\hertz}, \SI{0.88}{\volt}), Chimera reaches a peak performance of \SI{896}{\giga\ops} with a power consumption of \SI{600}{\milli\watt}, within the thermal power budget of passively cooled edge devices such as wearables or palm-sized robots.

\begin{table*}[t]
    \centering
    \caption{Comparison with State-of-the-Art (\gls{soa}).}
    \label{tab:soa}
    \scriptsize
    \begin{tabular}{lcccccc}
    \hline
     & \makecell{Ayaka JSSCC 24 \cite{ayaka_jssc24}} & \makecell{EVA VLSI 25 \cite{eva_vlsi25}} & \makecell{VLSI 25 \cite{fan_vlsi25}} & \makecell{TinyVers VLSI 22 \cite{tinyvers_jssc22}} & \makecell{ESSERC 24 \cite{stm_esserc24}} & \makecell{\textbf{Chimera} (This Work)} \\
     \hline
     & \multicolumn{3}{>{\columncolor{soaAccel}}c}{AI Accelerator} & \multicolumn{3}{>{\columncolor{soaSoc}}c}{\acrlong{ai-mcu}} \\
    Technology                        & \SI{28}{\nano\metre}              & \SI{16}{\nano\metre}              & \SI{22}{\nano\metre}              & \SI{22}{\nano\metre}                            & \SI{18}{\nano\metre}                    & \SI{22}{\nano\metre} \\
    Die Area                          & \SI{\ResultSoaAyakaDieArea}{\milli\metre\squared} & \SI{\ResultSoaEvaDieArea}{\milli\metre\squared} & \SI{\ResultSoaFanDieArea}{\milli\metre\squared} & \SI{\ResultSoaTinyVersDieArea}{\milli\metre\squared} & \SI{\ResultSoaStmDieArea}{\milli\metre\squared} & \best{\SI{\ResultSoaChimeraDieArea}{\milli\metre\squared}} \\
    Frequency                         & \SI{430}{\mega\hertz}             & \SI{1500}{\mega\hertz}            & \SI{400}{\mega\hertz}             & \SI{150}{\mega\hertz}                           & \SI{500}{\mega\hertz}                   & \SI{550}{\mega\hertz} \\
    Voltage                           & \SIrange{0.68}{1.0}{\volt}        & \SIrange{0.49}{0.9}{\volt}        & \SIrange{0.55}{0.9}{\volt}        & \SIrange{0.4}{0.9}{\volt}                       & \SIrange{0.5}{0.87}{\volt}              & \SIrange{0.6}{0.88}{\volt} \\
    \hline
    Application                       & Transformer        & Edge AI            & Transformer        & \makecell{IoT, DNN,\\ML, NSA}    & \makecell{Edge AI,\\IoT} & \makecell{\best{Edge AI GP,}\\\best{Transformer, ML}} \\
    Architecture                      & \makecell{Transformer\\Accelerator} & \makecell{1$\times$ RISC-V +\\PE Array} & \makecell{Transformer\\Accelerator} & \makecell{1$\times$ RI5CY +\\ML Accel.} & \makecell{1$\times$ RISC-V +\\128 PEs TPU} & \makecell{\best{1$\times$ RV32IMC +}\\\best{9$\times$ RV32IMA +}\\\best{1$\times$ TAC Accel.}} \\
    GP-Host / Prog. Accel.            & \xmark{} / \xmark  & \xmark{} / \xmark  & \xmark{} / \xmark  & \cmark{} / \xmark                  & \cmark{} / \xmark          & \best{\cmark{} / \cmark} \\
    L2 Aggr. Bandwidth                & --               & --               & --               & --                               & --                       & \best{\SI{563}{\giga\bs}} \\
    Precision                         & \makecell{INT16/8} & FP16 & INT4/8        & \makecell{INT2/4/8}     & INT8                       & INT8 \\
    \hline
    Peak Performance                         & \makecell{\num{\ResultSoaAyakaPeakPerfMin}--\textbf{\num{\ResultSoaAyakaPeakPerfMax}}\textsuperscript{*}~\si{\giga\ops}\\(INT8, @\SI{430}{\mega\hertz})} & \makecell{\SI{\ResultSoaEvaPeakPerf}{\giga\ops}\\(FP16, @\SI{1.5}{\giga\hertz})} & \makecell{\SI{\ResultSoaFanPeakPerf}{\giga\ops}\\(INT8, @\SI{400}{\mega\hertz})} & \makecell{\SI{\ResultSoaTinyVersPeakPerf}{\giga\ops}\\(INT8, @\SI{150}{\mega\hertz})} & --                 & \makecell{\SI{\ResultSoaChimeraPeakPerf}{\giga\ops}\\(INT8, @\SI{550}{\mega\hertz})} \\
    Peak Efficiency               & \makecell{\num{\ResultSoaAyakaPeakEffMin}--\textbf{\num{\ResultSoaAyakaPeakEffMax}}\textsuperscript{*}~\si{\tera\ops/\watt}\\(@\SI{0.68}{\volt}, INT8)} & \makecell{\SI{\ResultSoaEvaPeakEff}{\tera\ops/\watt}\\(@\SI{0.49}{\volt}, FP16)} & \makecell{\num{\ResultSoaFanPeakEffMin}--\num{\ResultSoaFanPeakEffMax}\textsuperscript{\ddag}~\si{\tera\ops/\watt}\\(@\SI{0.55}{\volt}, INT8)} & \makecell{\SI{\ResultSoaTinyVersPeakEff}{\tera\ops/\watt}\\(@\SI{0.4}{\volt}, INT8)} & \makecell{\SI{\ResultSoaStmPeakEff}{\tera\ops/\watt}\textsuperscript{\dag}\\(@\SI{0.5}{\volt}, INT8)} & \makecell{\best{\SI{\ResultSoaChimeraPeakEff}{\tera\ops/\watt}}\\(@\SI{0.6}{\volt}, INT8)} \\
    Area Efficiency                      & \num{\ResultSoaAyakaAreaEffMin}--\textbf{\num{\ResultSoaAyakaAreaEffMax}}\textsuperscript{*}~\si{\giga\ops/\milli\metre\squared} & \SI{\ResultSoaEvaAreaEff}{\giga\ops/\milli\metre\squared} & \SI{\ResultSoaFanAreaEff}{\giga\ops/\milli\metre\squared} & \SI{\ResultSoaTinyVersAreaEff}{\giga\ops/\milli\metre\squared} & -- & \best{\SI{\ResultSoaChimeraAreaEff}{\giga\ops/\milli\metre\squared}} \\
    \hline
    \multicolumn{7}{l}{\textsuperscript{*}The highest values are measured assuming 90\% output sparsity. \textsuperscript{\ddag}Value obtained with one dense and one 87.5\% sparse input.} \\
    \multicolumn{7}{l}{\textsuperscript{\dag} Scaling from 0.5 to \SI{0.6}{\volt}, the efficiency is \ResultSoaStmPeakEffScaledDeficit\% lower than Chimera.} \\
    \end{tabular}
    \vspace{-0.7em}
\end{table*}

\begin{table}[t]
    \centering
    \caption{Full network evaluation derived from silicon measurements.}
    \vspace{-0.3em}
    \label{tab:network}
    \begin{tabular}{lccc}
    \hline
     & MobileBERT & \makecell{Whisper-Tiny\\Encoder} & DINOv2-S \\
    \hline
    Model Complexity [GOP] & 7.4     & 9.7       & 11.7 \\
    Throughput [1/s]     & 7.7--21 & 2.0--5.4  & 1.2--3.3 \\
    Energy [\SI{}{\milli\joule}]        & 9.2--16 & 36--72    & 60--118 \\
    \hline
    \end{tabular}
    \vspace{-1em}
\end{table}

In \Cref{tab:soa}, we compare Chimera with both \gls{soa} transformer accelerators and \glspl{ai-mcu} in comparable technology nodes.
Compared to full \glspl{mcu} architectures, Chimera achieves \ResultSoaChimeraPeakEffRatioSoc$\times$ higher energy efficiency and \ResultSoaChimeraAreaEffRatioSoc$\times$ higher area efficiency.
Compared to pure accelerators, our architecture still achieves competitive energy efficiency for dense workloads, while providing significantly greater flexibility thanks to the tightly integrated RV processors and L2 memory hierarchy support.
In addition, it achieves up to \ResultSoaChimeraAreaEffRatioAcc$\times$ higher area efficiency.
Unlike prior \glspl{ai-mcu} that primarily target CNN workloads \cite{tinyvers_jssc22,stm_esserc24}, this work evaluates more advanced transformer workloads in \Cref{tab:network}, with energy per inference of 9.2/36/60 \SI{}{\milli\joule} for MobileBERT, Whisper-Tiny Encoder, and DINOv2-S, achieving up to \SI{737}{\giga\ops} and \SI{2.54}{\tera\ops/\watt}.

\section*{Acknowledgment}
\ifdefined\blindreview
\textit{Omitted for blind review.}
\else
This work is funded in part by the Convolve project evaluated by the EU Horizon Europe research and innovation programme under grant agreement No. 101070374 and has been supported by the Swiss State Secretariat for Education Research and Innovation under contract number 22.00150.
\fi

\bibliography{paper}
\bibliographystyle{IEEEtran}

\end{document}